\documentstyle[aps,epsf,preprint]{revtex}
\begin{document}

\baselineskip=24 pt

%\twocolumn[\hsize\textwidth\columnwidth\hsize\csname@twocolumnfalse\endcsname

\title
{Tricritical  transition in the classical $XY$ model on Kagom\'e
lattice under local anisotropy }
\author
{Farhad Shahbazi {\footnote {Electronic address:
shahbazi@cc.iut.ac.ir }}, Saba Mortezapour  {\footnote {Electronic
address: mortezapour-s@ph.iut.ac.ir }}}
\address
{\it Dept. of Physics , Isfahan University of Technology,
84156-83111, Isfahan, Iran. \\}

\maketitle

\begin{abstract}
Using  mean-field theory and  high resolution  Monte Carlo
simulation technique based on multi-histogram method, we have
investigated the critical properties of an antiferromagnetic $XY$
model on the 2D Kagom\'e lattice, with single ion easy-axes
anisotropy. The mean-field theory predicts second-order phase
transition from disordered to all-in all-out state for any value
of anisotropy for this model. However, Monte Carlo simulations
result in first order  transition for small values of anisotropy
which turns to second order with increasing  strength of
anisotropy, indicating the existence of a tricritical point for
this model. The critical exponents, obtained by finite-size
scaling methods, show that the transition is  in  Ising
universality class for large values of anisotropy, while the
critical behaviour of the system deviates from 2D-$\phi^6$ model
near the tricritical point. This suggests the possibility for
existence of a new tricritical universality in two-dimensions.
\end{abstract}
\hspace{.3in}
\newpage

\section{introduction}

The phenomenon of geometric frustration has attracted the
interest of physicists due to the presence of degeneracy in the
classical ground states arising from the arrangement  of spins on
triangular clusters \cite{greedan,ramirez1,ramirez2,moessner}. A
frustrated magnet is one in which not all interaction energies
can be simultaneously optimized, for which the anti-ferromagnetic
Ising model on a two-dimensional triangular lattice, is an
example. The highly frustrated magnets, on the other hand, are
the class of frustrated magnets that have an infinite number of
classical ground states, even after removing the global
symmetries of Hamiltonian.

The classical $XY$ anti-ferromagnet on the two-dimensional
Kagom\'e lattice constructed from corner-sharing triangular units
and the classical Heisenberg antiferromagnet on the $3D$
pyrochlore lattice consisting of corner-sharing tetrahedra are two
prototypes  of the highly frustrated class. The discoveries, such
as heavy-fermion behaviour \cite{heavy-fermion}, spin-ice
ordering \cite{ice1,ice2,ice3}, spin nematics \cite{nematic},
spin liquid behaviours \cite{liquid1,liquid2,liquid3} and even
novel superconductivity \cite{sc} in materials with magnetic
sublattices of corner-sharing tetrahedra (such as spinel and
pyrochlores), have made these structures in the focus of
physicists's attention over the recent years.

It has been widely accepted  that no order-by-disorder mechanism
can  establish a long-range order in the Heisenberg pyrochlore
anti-ferromagnet, consequently such a system remains disordered
at all temperatures~\cite{od1,od2,od3}. However, experimental
observations have represented an all-in all-out long-range order
(consisting of four sublattices oriented along four [111] spin
directions), for the low-temperature phase of $\mathrm{FeF_{3}}$
in pyrochlore form~\cite{exp1,exp2}. In this compound, the
$\mathrm{Fe^{+3}}$ ions located on a pyrochlore lattice, interact
anti-ferromagnetically with their nearest neighbors. Since, the
magnetic {$\mathrm{Fe}^{+3}$} ions are in $d^{5}$ electronic
configuration with a totally symmetric ground state and no net
angular momentum, this system can be considered as  a Heisenberg
anti-ferromagnet and so the origin of the long-range ordered
phase in it, has remained  as a puzzle. Reimers \textit{et al}
have shown that, taking into account  the interaction with
farther neighbors, would cause a second order transition in this a
system~\cite{Reimers}. However, they found that because of the
thermal fluctuations, a co-linear spin ordering would be preferred
rather than the all-in all-out state. Therefore, it seems that to
stabilize a long range all-in all-out spin configuration, one
should inevitably introduce  a single-ion an-isotropic crystal
field  term in the model Hamiltonian. Another interesting aspect
of the transition in pyr-$\mathrm{Fe^{+3}}$ is in its
universality class. The  order parameter critical exponent
$\beta$ has been fixed to the value $0.18(2)$, in neutron-
diffraction experiments, which is nearest to  the tetra-critical
value $\beta=1/6$~\cite{sim}. On the other hand, recent Monte
Carlo simulations, carried on Heisenberg pyrochlore
antiferromagnet with single ion anisotropy, have revealed the
existence of a tricritical point for this system
\cite{peter,kawamura}.

The above interesting problem motivated us to study the critical
properties of its two-dimensional equivalent, the $XY$ Kagom\'e
anti-ferromagnet  model with single-ion anisotropy. The classical
antiferromagnetic $O(n)$ models on the Kagom\'e lattice have been
studied by Huse and Rutenberg ~\cite{order1}. There, it has been
shown that the Ising model ($n=1$) is disordered at all
temperatures, while the $XY$ model ($n=2$) represents quasi
long-range order in a three-fold ordered parameter at zero
temperature. Because the system is two-dimensional this quasi
long-range order does not survive at finite temperatures and so
transforms to disordered phase through a Kosterlitz-Thouless
transition. The ground state of the $XY$ model has the same
properties as the three-state Potts model which  can be mapped
exactly onto solid-on solid (SOS) model at the roughening
transition. On the other hand, the study of two-dimensional
antiferromagnet Heisenberg model on Kagom\'e, have  been carried
out by Ritchey \textit{et al}, which resulted in a coplanar spin
configurations in which there are nematic spin correlations with
planar threefold symmetry and non-Abelian homotopy
~\cite{chandra}. They have also shown that very small amounts of
bond  XY anisotropy are sufficient to convert a crossover to a
topological phase transition, in which the binding of non-Abelian
disclinations would result in a glassy behavior in the absence of
extrinsic disorder.

The Hamiltonian of nearest-neighbor $XY$ antiferromagnet model on
the $\mathrm{Kagom\acute{e}}$ lattice is given by:

\begin{equation}\label{H1}
H=-J\sum_{\langle ij\rangle}{\bf S}_{i}.{\bf  S}_{j},
\end{equation}
in which $J\langle 0$ and ${\bf S}_{i}$ denotes the unit planar
vectors and $\langle ij\rangle$ indicates the nearest-neighbors.
The ground state of this model   is known to have a huge
accidental degeneracy not related to the global symmetries of the
Hamiltonian \cite{order1,order2}. In any ground state of the
$\mathrm{Kagom\acute{e}}$ lattice the spins ${\bf S}_{i}$
acquires only three directions whose angles with respect to an
arbitrary axis, say $x$-axis, differ from each other by $2\pi/3$.
Therefore, the ground state in addition to the continuous $U(1)$
symmetry (due to the arbitrary simultaneous rotation of all
spins) is characterized by a well developed discrete degeneracy
of the same type as in the 3-state antiferromagnetic Potts model.

The extensive degeneracy of the ground state in this model makes
it extremely unstable towards the imposing of perturbations
\cite{pyro}. For instance, if one adds a single-ion easy-axis
anisotropic term to  Hamiltonian (\ref{H1}), all spins prefer to
align along the anisotropy directions  yielding a long-range
all-in all-out state for the system.

The goal of this  paper is to determine  the critical properties
of an $XY$ model  on the two dimensional Kagom\'e lattice  with
single ion easy-axes anisotropic term. For this purpose we employ
mean-field theory and Monte Carlo simulation.

The structure of paper is as follows. In Sec. II, we introduce a
mean-field formalism to derive the qualitative picture of
transitions in the model.  Section III is dedicated to the Monte
Carlo method based on multiple histograms and also some methods
for analyzing the Monte Carlo data to determine the order of
transitions, critical temperatures and critical exponents. The
simulation results and discussion are given in Sec. IV and
conclusion  appears in Sec. V.

\section{mean-field formalism}

The Hamiltonian, describing the $XY$ spins with nearest-neighbor
anti-ferromagnetic interaction on a Kagom\'e lattice subjected to
single site easy-axes anisotropy, is given by :

\begin{equation}\label{H}
H=-{J\over 2}\sum_{i,j}\sum_{a,b}{\bf S}_{i}^{a}\cdot{\bf
S}_{j}^{b}-D\sum_{i}\sum_{a}({\bf S}_{i}^{a}\cdot\hat{\mathrm{
z}}^{a})^{2},
\end{equation}
in which $J < 0, D > 0$ and $i,j=1,\cdot\cdot\cdot ,N$ and
$a,b=1,2,3$ denote the Bravais lattice and sublattice indices,
respectively. $\hat{\mathrm{z}}^a$'s represent the  unit vectors
of three easy-axes directions in 2d plane, which are along the
line connecting the corner and the center of corner-sharing
triangular units, given by:

\begin{eqnarray}
\hat{\mathrm{z}}^{1}&=&({\sqrt{3}\over 2},{-1\over 2})\nonumber\\
\hat{\mathrm{z}}^{2}&=&(-{\sqrt{3}\over 2},{-1\over 2})\nonumber\\
\hat{\mathrm{z}}^{3}&=&(0,1)
\end{eqnarray}
in global Cartesian coordinates.

To apply mean-field theory on this model, we follow the method
introduced by Harris, Mouritson and Berlinsky
\cite{Reimers,Harris}. Defining the average magnetization as
$\bf{M}_{i}^{a}=\langle\bf{S}_{i}^{a}\rangle$ and the deviation
from the mean magnetization as $\delta{\bf
S}_{i}^{a}=\bf{S}_{i}^{a}-\bf{M}_{i}^{a}$, to order $O(\delta
S^2)$, we can write the Hamiltonian (Eq.(\ref{H})) as the
following linear form:

\begin{equation}\label{H-mf}
H={J\over 2}\sum_{i,j}\sum_{a,b}{\bf M}_{i}^{a}\cdot{\bf
M}_{j}^{b}+D\sum_{i}\sum_{a}({\bf
M}_{i}^{a}\cdot\hat{\mathrm{z}}^{a})^{2}
-{J}\sum_{i,a}\sum_{j,b}{\bf M}_{j}^{b}\cdot{\bf
S}_{i}^{a}-2D\sum_{i,a}({\bf
S}_{i}^{a}\cdot\hat{\mathrm{z}}^{a})({\bf
M}_{i}^{a}\cdot\hat{\mathrm{z}}^{a}).
\end{equation}

Therefore, the mean-field  partition function can be written as:

\begin{equation}\label{pt}
Z=e^{-\beta\left({J\over 2}\sum_{i,j}\sum_{a,b}{\bf
M}_{i}^{a}\cdot{\bf M}_{j}^{b}+D\sum_{i}\sum_{a}({\bf
M}_{i}^{a}\cdot\hat{\mathrm{z}}^{a})^{2}\right)} \Pi_{i,a}\int
e^{\beta{\bf B}_{i}^{a}\cdot{\bf S}_{i}^{a}}d{\bf S}_{i}^{a},
\end{equation}
where
\begin{equation}
{\bf B}_{i}^{a}={J}\sum_{j\neq i}\sum_{b\neq a}{\bf
M}_{j}^{b}+2D({\bf M}_{i}^{a}\cdot\hat{\mathrm{z}}^{a})\hat
{\mathrm{z}}^{a},
\end{equation}
in which, the summation is over the nearest neighbors. The
integral in Eq.(\ref{pt}) can be evaluated easily as follows:

\begin{equation}
\int e^{\beta{\bf B}_{i}^{a}\cdot{\bf S}_{i}^{a}}d{\bf S}_{i}^{a}=
2\pi\int_{0}^{\pi} e^{\beta{B}_{i}^{a}{\cos(\theta)}}d\theta=2\pi
I_{0}(\beta B_{i}^{a}) ,
\end{equation}
where $B_{i}^{a}=|{\bf B}_{i}^{a}|$. Then, assuming $K_{B}=1$, we
reach the following expression for the free energy:

\begin{equation}\label{f}
F=-T\ln{Z}={J\over 2}\sum_{i,j}\sum_{a,b}{\bf M}_{i}^{a}\cdot{\bf
M}_{j}^{b}+D\sum_{i}\sum_{a}({\bf
M}_{i}^{a}\cdot\hat{\mathrm{z}}^{a})^{2}-T\sum_{i,a}\ln\left(2\pi
I_{0}({ B_{i}^{a}\over T}) \right).
\end{equation}

From the mean-field free energy, obtained above, one can calculate
the magnetization and entropy as:

\begin{eqnarray}
S&=&-{\partial F\over \partial T}=\sum_{i,a}\ln\left(2\pi I_{0}({
B_{i}^{a}\over T})\right)+{1\over T^2}\sum_{i,a}
{B_{i}^{a}I_{1}({B_{i}^{a}\over T})\over 2I_{0}({ B_{i}^{a}\over
T})}\\, {\bf M}_{i}^{a}&=&-\nabla_{B} F=-{\partial F\over
\partial B_{i}^{a}}{\hat B_{i}^{a}}=-\sum_{i,a} {I_{1}
({B_{i}^{a}\over T})\over 2I_{0}({B_{i}^{a}\over T})}.
\end{eqnarray}

For small values of $B$, one can expand Eq.(10) as:

\begin{equation}
{M}_{i}^{a}=\left[{B_{i}^{a}\over 2T} - {{B_{i}^{a}}^{3}\over
16T^3}+{{B_{i}^{a}}^{5}\over 96T^5}-{11 \over 6144}
{{B_{i}^{a}}^{7}\over T^7}+ O({{B}^{9}})\right],
\end{equation}
from which, by reversing the series one gets:

\begin{equation}\label{exp-b}
{B}_{i}^{a}={2T}{{M_{i}^{a}}}-T{({M_{i}^{a}})^{3}}+{5\over
9}{({M_{i}^{a}})^5} +O(M^8).
\end{equation}

Substituting Eq.(\ref{exp-b}) into Eq.(9) and expanding the
entropy in powers of $M_{i}^{a}$, enables us to expand the free
energy as:

\begin{eqnarray}
F&=&\langle H \rangle -TS\nonumber \\
&=& -4NT\ln(4\pi)-{J\over 2}\sum_{i,j}\sum_{a,b}{\bf
M}_{i}^{a}\cdot{\bf M}_{j}^{b}-D\sum_{i,a}({\bf
M}_{i}^{a}\cdot\hat{\mathrm{z}}^{a})^{2}\nonumber \\
&+&T\sum_{i,a}\left(({M_{i}^{a}})^{2}+{1\over
4}({M_{i}^{a}})^{4}-{5\over 36}({M_{i}^{a}})^{6}+O(M^7)\right),
\end{eqnarray}
where we have used Eq.(\ref{H-mf}). We can also expand the free
energy in terms of Fourier components defined by:

\begin{eqnarray}
{\bf M}_{i}^{a}&=&\sum_{q} {\bf M}_{\bf q}^{a} \exp(i{\bf
q}\cdot{\bf R}_{i}^{a})\\
 J_{\bf q}^{ab}&=&\sum_{j\neq
i}\sum_{b\neq a} {J} \exp\left(i{\bf q}\cdot({\bf R}_{i}^{a}-{\bf
R}_{j}^{b})\right),
\end{eqnarray}
where the summation in Eq.(15) is over the nearest neighbors of a
selected spins. Then we reach the following form for the free
energy per particle in terms of Fourier components:

\begin{eqnarray}\label{ff2}
f(T,J,D)&=&\frac{F(T,J,D}{N}=-4T\ln(4\pi)\nonumber \\
&+&{1\over 2}\sum_{q}\sum_{ab} {\bf M}_{\bf q}^{a} {\bf M}_{-\bf
q}^{b}(2T\delta^{ab}-J_{\bf q}^{ab})-D\sum_{q}\sum_{a}({\bf M_{\bf
q}^a}\cdot{\hat z}^{a})({\bf M_{-\bf q}^a}\cdot{\hat
z}^{a})\nonumber \\
&+&{1\over 4}T\sum_{a}\sum'_{\{{\bf q}\}}({\bf M}_{\bf
q1}^{a}\cdot{\bf M}_{\bf q2}^{a})({\bf M}_{\bf q3}^{a}\cdot{\bf
M}_{\bf q4}^{a}) \nonumber \\
&-&{5\over 36}T\sum_{a}\sum'_{\{{\bf q}\}}({\bf M}_{\bf
q1}^{a}\cdot{\bf M}_{\bf q2}^{a})({\bf M}_{\bf q3}^{a}\cdot{\bf
M}_{\bf q4}^{a})({\bf M}_{\bf q5}^{a}\cdot{\bf M}_{\bf
q6}^{a})+O(M^7),
\end{eqnarray}
where

\[
\sum'_{\{{\bf q}\}}=\sum_{\{{\bf q}\}}\delta(\sum_{i}{\bf qi}).
\]

The free energy (Eq.\ref{ff2}) can be rewritten in terms of
Cartesian components of \\
${\bf M}_{\bf q}^{a}=({m}_{\bf q}^{a,1},{m}_{\bf q}^{a,2})$ as:

\begin{eqnarray}\label{f2}
f(T,J,D)&=&-4T\ln(4\pi)+{1\over
2}\sum_{q}\sum_{ab}\sum_{\alpha\beta}
(2T\delta^{ab}\delta^{\alpha\beta}-J_{\bf
q}^{ab}\delta^{\alpha\beta}-D_{\alpha\beta}^{a}\delta^{ab}){m}_{\bf q}^{a,\alpha}{m}_{-\bf q}^{b,\beta}\nonumber \\
&+&{1\over 4}T\sum_{a}\sum_{\alpha\beta}\sum'_{\{{\bf
q}\}}({m}_{\bf q1}^{a,\alpha}{m}_{\bf q2}^{a,\alpha})({m}_{\bf
q3}^{a,\beta}{m}_{\bf q4}^{a,\beta}) \nonumber \\
&-&{5\over 36}T\sum_{a}\sum_{\alpha\beta\gamma}\sum'_{\{{\bf
q}\}}({m}_{\bf q1}^{a,\alpha}{m}_{\bf q2}^{a,\alpha})({m}_{\bf
q3}^{a,\beta}{m}_{\bf q4}^{a,\beta})({m}_{\bf
q5}^{a,\gamma}{m}_{\bf q6}^{a,\gamma})+O(M^7),
\end{eqnarray}
in which $\alpha,\beta,\gamma$ take the values $1,2$.  It can be
seen from the above equation, that only the an-isotropic term $D$
couples the different Cartesian components of ${\bf M}$. The
$2\times 2$ matrices $D^{a}$ are given by:

\begin{equation}
 D^{1}= D\left(
          \begin{array}{cc}
              {\sqrt{3}\over2} & {\sqrt{3}\over4}\\
              {\sqrt{3}\over4} & {1\over2}

          \end{array} \right),
 D^{2}= D\left(
         \begin{array}{cc}
              {\sqrt{3}\over2} & -{\sqrt{3}\over4} \\
              -{\sqrt{3}\over4} & {1\over2}

         \end{array} \right),
%\end{equation}
%\begin{equation}
D^{3}= D\left(
         \begin{array}{cc}
              {0} & {0} \\
              {0} & {1}

         \end{array} \right).
\end{equation}

Thus we are left with the following coupling $6\times 6$ matrix
for the quadratic terms:

\begin{equation}\label{jq}
 {\tilde J}_{\bf q}= D\left(
         \begin{array}{ccc}
               D^{1} & J_{\bf q}^{12}& J_{\bf q}^{13}\\
               J_{\bf q}^{12} & D^{2}& J_{\bf q}^{23}\\
               J_{\bf q}^{13} & J_{\bf q}^{23}& D^{3}

          \end{array} \right),
\end{equation}
in which the off-diagonal matrices $J_{\bf q}^{ij}$ are
proportional to the $2\times 2$ unit matrix as follows :

\begin{eqnarray}
J_{\bf q}^{12}&=&2J\cos(\frac{q_{x}}{2})I_{2\times 2}\\
J_{\bf q}^{13}&=&2J\cos(\frac{\sqrt{3}q_{y}+q_{x}}{2})I_{2\times 2}\\
J_{\bf q}^{23}&=&2J\cos(\frac{\sqrt{3}q_{y}-q_{x}}{2})I_{2\times
2}.
\end{eqnarray}

In deriving the above expressions, we have used Eq.(15) together
with the positions of $\mathrm{Kagom\acute{e}}$ atoms given by
their $xy$ components. For convenience we  reduce the number of
indices ($a=1,2,3$ and $\alpha=1,2$) by defining a new set of
indices $s=1,\cdot\cdot\cdot,6$, which leads to a $6$-component
magnetization vector as:

\begin{equation}
{\tilde{\bf M}}_{\bf q}=(m_{\bf q}^{1,1},m_{\bf q}^{1,2},m_{\bf
q}^{2,1},\cdot\cdot\cdot m_{\bf q}^{3,2}) =(m_{\bf q}^{1},m_{\bf
q}^{2},\cdot\cdot\cdot m_{\bf q}^{6}),
\end{equation}
from which the quadratic term in free energy can be written as:

\begin{equation}
f^{(2)}=\sum_{\bf q}{\tilde{\bf M}}_{\bf q}.{\tilde
J}_{q}.{\tilde{\bf M}}_{\bf q}^{T}.
\end{equation}

Diagonalizing the quadratic term, requires transforming to the
normal modes $\Phi_{\bf q}$:

\begin{equation}
m_{\bf q}^{s}=\sum_{i=1}^{6} U_{\bf q}^{si} \phi_{\bf q}^{j}
\end{equation}
for $s=1,2, \cdot\cdot\cdot 6$.

$U_{\bf q}$ is the unitary matrix that diagonalizes the coupling
matrix ${\tilde J}_{\bf q}$, with eigenvalues $\lambda_{\bf
q}^{i}$:

\begin{equation}\label{normal}
 \sum_{b} {\tilde J}_{\bf q}^{ab}U_{\bf q}^{bi}=\lambda_{\bf q}^{i}U_{\bf q}^{ai}.
\end{equation}
in which, the unitarity condition requires:

\begin{equation}\label{unitary}
\sum_{a} U_{\bf q}^{ai} U_{-\bf q}^{aj}=\delta^{ij}.
\end{equation}

Equation (\ref{normal}) enables  us to write the free energy as a
power series in terms of normal modes, such that to $O(\phi^7)$ we
obtain the following expansion for the free energy:

\begin{eqnarray}\label{f3}
f(T,J,D)&=&-4T\ln(4\pi)+{1\over 2}\sum_{q}\sum_{i=1}^{12} (2T-\lambda_{\bf q}^{i}){\phi}_{\bf q}^{i}\phi_{-\bf q}^{i}\nonumber \\
&+&{T\over 4} \sum_{s=1}^{12}\sum_{ijkl}\sum'_{\{{\bf q}\}}
U_{\bf q1}^{si} U_{\bf q2}^{sj}U_{\bf q3}^{sk}U_{\bf q4}^{sl}
{\phi}_{\bf q1}^{i}{\phi}_{\bf q2}^{j}{\phi}_{\bf q3}^{k}{\phi}_{\bf q4}^{l} \nonumber \\
&+&{5\over 36} T\sum_{s=1}^{12}\sum_{ijklmn}\sum'_{\{{\bf q}\}}
U_{\bf q1}^{si} U_{\bf q2}^{sj}U_{\bf q3}^{sk}U_{\bf
q4}^{sl}U_{\bf q5}^{sm}U_{\bf q6}^{sn} {\phi}_{\bf
q1}^{i}{\phi}_{\bf q2}^{j}{\phi}_{\bf q3}^{k}{\phi}_{\bf q4}^{l}
{\phi}_{\bf q5}^{m} {\phi}_{\bf q6}^{n}.
\end{eqnarray}

It is clear that phase transition occurs when the sign of
quadratic term  of free energy changes. Therefore, from the above
expression one finds that the spontaneously breaking symmetry
occurs  at a temperature:

\begin{equation}
T_{c}={1\over 2}{\bf max}_{{\bf q},i}\{{\lambda_{\bf q}^{i}}\},
\end{equation}
where max $\{ \}$ means the global maximum over all $i$ and ${\bf
q}$. In the  case of $D=0$ one can exactly diagonalize the matrix
${\tilde J}_{\bf q}$ (Eq.(\ref{jq})) and find the following
eigenvalues:

\begin{eqnarray}
\lambda_{\bf q}^{i}&=&-2J  \hspace{3.1cm} i=1,2 \nonumber\\
\lambda^{i}_{\bf q}&=&2J(1-\sqrt{3+Q}) \hspace{1cm} i=3,4 \nonumber\\
\lambda^{i}_{\bf q}&=&2J(1+\sqrt{3+Q}) \hspace{1cm} i=5,6,
\end{eqnarray}
where $Q$ is given by:

\begin{eqnarray}
Q=
\{&&\cos(2q_{x})+\cos(\sqrt{3}q_{x}+q_{y})+\cos((2-\sqrt{3})q_{x}-q_{y})\},
\end{eqnarray}
which coincides with the result derived in  Ref.\cite{Reimers}.
The above results show that for $J < 0$ the largest eigenvalues
are degenerate and dispersionless (q-independent), such that when
$T < -J$, the order parameters corresponding to all of these
modes turn to be nonzero and we were left with a huge number of
states with  broken symmetry. Therefore, because of the extensive
degeneracy of symmetry broken states, one concludes that in
mean-field theory, no long range order can be established as the
temperature decreases down to zero. The $q$-dependence of
eigenvalues for $D=0$ along [1~0] direction is depicted in
Fig.(1).

For an-isotropic case ($D \rangle 0$) the eigenvalues of matrix
${\tilde J}_{\bf q}$ can  be obtained numerically. The dispersion
curves for $D=0.2$ and $D=1.0$ along [1~0] direction has been
shown in Figs.(2) and (3), respectively. As can be seen from these
graphs, all the degeneracies have been removed, so we were left
with 6 distinct modes, where  the highest mode has a maximum at
$q=0$ with the value $\lambda_{0}^{1}=-2J+2D$. It can be easily
shown, by deriving the eigenvector of this  mode, that this mode
corresponds to all-in all-out spin configuration represented in
Fig.(4). As a result, the mean-field theory predicts a continues
phase transition from disordered to a long-range ordered all-in
all-out state at the critical temperature $T_{c}=-J+D$. Another
interesting point is that the branch $\lambda_{\bf q}^{4}$ is
independent of magnitude of anisotropy ($D$), which means that
the modes describing by it, are corresponding to spin fluctuations
perpendicular to easy-axes directions (${\hat{
\mathrm{z}}}^{a},a=1,2,3$) in Hamiltonian, given by Eq.(\ref{H}).

\section{Monte Carlo simulation}

 For large values of $D$, spins tend
to remain mainly along easy-axes directions such that the
effective degrees of freedom flip along these axes. Therefore, one
expects that the transition to all-in all-out state to be in 2D
Ising universality class. However, when $D$ is small, the
transverse fluctuations normal to local easy-axes directions
become larger and so this leads to lowering of  the transition
temperature as well as deviation from Ising behaviour. In this
section we use Monte Carlo simulation, to study the phase
transition of the model described in previous section and find the
order of transitions for different values of anisotropic term $D$.

To obtain a qualitative picture of the transitions and also the
approximate location of the critical points, we first set some
low resolution simulations. The simulations were carried out using
standard Metropolis single spin-rotating algorithm with lattice
size $N=3\times 20 \times 20$. During each simulation step, the
angles of planar spins with the horizontal axes were treated as
unconstrained, continuous variables. The random-angles rotations
were adjusted in such a way that roughly $50\%$ of the attempted
angle rotations were accepted. To ensure thermal equilibrium, 100
000 Monte Carlo steps (MCSs) per spin were used for each
temperature and 200 000 MCS were used for data collection. The
basic thermodynamic quantities of interest are the specific heat
$c=(\langle E^2 \rangle-\langle E \rangle^{2})/(N T^{2})$, the
order parameter defined as $M=|\sum_{i,a}{\bf
S}_{i}^{a}\cdot\hat{\mathrm{z}}^{a}|/N$ and the susceptibility
$\chi=(\langle M^2 \rangle-\langle M \rangle^{2})/(NT)$.

In Figs. (5-8), temperature dependence of the energy per spin,
,the order parameter, specific-heat and  susceptibility  have
respectively been represented for $J=-1.0$, $D=0.2,0.1$. As can
be observed from  Figures.(7) and (8),  the transition for $D=0.2$
seems to be continuous, while for $D=0.1$, because of sudden peaks
in specific heat and susceptibility, it  seems to be first order.
However, The determination of the order of transition requires
more accurate methods, for which we will use Binder's fourth
energy cumulant method. Once the probability density of energy
($P(E,T)$) is obtained, for measuring the thermodynamic
quantities other than the energy, one can choose to work with
this energy probability distribution and microcanonical averages
of the quantities of interest. This leads to optimized use of
computer memory. The microcanonical average of a given quantity
$A$, which is a function of energy, can be calculated directly as:

\begin{equation}
A(E)=\frac{\sum_{t}A_{t}\delta_{E_{t},E}}{\sum_{t}\delta_{E_{t},E}},
\end{equation}
from which, the canonical average of $A$ can be obtained as a
function of $T$:

\begin{equation}
\langle A \rangle=\frac{\sum_{E}A(E)P(E,T)}{\sum_{E}P(E,T)}.
\end{equation}

In our simulation, we use $\mathrm{Kagom\acute{e}}$ lattices with
linear sizes $L=20,24,28,32,36,40$ (the number of sites is given
by $N=3\times L\times L$), such that the maximum number of spins
is 4800, large enough for reducing the finite size effects. For
each  system size, at least five overlapping energy histograms
are obtained near the transition point so that the statistical
uncertainty in the wing of the histograms, may be suppressed by
using the optimized multiple-histogram method\cite{fs}. This
enables us to measure the location and magnitude of the extrema
of the thermodynamic quantities with high accuracy. For each
histogram we performed $5\times10^5$ Monte Carlo steps per spin
for equilibration  and also  $5\times10^5$ MCSs for gathering
data. To reduce the correlation, 10 to 20 Monte Carlo sweeps were
discarded between successive measurements. In all simulation we
fix $J=-1$ and vary the value of $D$ from 0.1 to 1.0. First of
all, we deal with the order of transitions.

\subsection{Order of the transition}

To determine the order of transitions, we used Binder's fourth
energy cumulant defined as:

\begin{equation}
 U_{L}=1-\frac{<E^4>}{3<E^2>^2}.
\end{equation}
It has been shown that this quantity reaches a minimum at the
effective transition temperature $T_{c}(L)$ whose  size
dependence is given by\cite{landau,lk,lb}:

\begin{equation}\label{bind}
U_{min}(L)=U^{*}+BL^{-d}+O(L^{-2d}),
\end{equation}
where

\begin{equation}
U^{*}=\frac{2}{3}-\left(e_{1}/e_{2}-e_{2}/e_{1}\right)^{2}/12.
\end{equation}
The quantities $e_{1}$ and $e_{2}$ are the values of energy per
site at the transition point of a first order phase transition and
$d$ is the spatial dimension of the system ($d=2$ in our
simulation). Hence, for the continuous transitions for which
there is no latent heat ($e_{1}=e_{2}$), in the limit of infinite
system sizes, $U_{min}(L)$ tends to the value $U^{*}$ equal to
$2/3$. For the first-order transitions, however $e_{1}\neq e_{2}$
and then $U^{*}$ reaches a value less than $2/3$ in the the limit
$L\rightarrow\infty$.

The size dependences of ${U(L)}$ for $D=1.0,0.18,0.15,0.13,0.1$
have been exhibited in Fig.(9). The straight lines fitted to the
data have been obtained from Eq.(\ref{bind}). The values of
$U^{*}$ and latent heat per spin are also listed in Table.(I),
from which one can see that, within the errors of simulation,
transitions are second order for $D>0.17$ and
 clearly first order for $D<0.15$. The precise  determination of the tricritical point is
 extremely difficult, however our results suggest the
existence of a tricritical point between $D/|J|=0.15$ and
$D/|J|=0.17$.

In the Figs.(10) and (11) the energy histograms of $D=0.2$ and
$D=0.1$ for the size $N=3\times 40\times 40$ have been shown,
respectively. As can be seen from these figures, the energy
histogram for $D=0.2$ has one broad peak at the transition, while
for $D=0.1$, it has two well separated peaks around the
transition temperature. This is in agreement with the results of
Binder's method. Note that the small peak at the middle in
Fig.(11), is artifact of the finite time of simulation and will
vanish at large enough times. The reason is that at a  strong
first order transition point,free energy  possesses two
equivalent minima  corresponding to two stable coexisting phases.
For large system sizes these two minima are separated by a large
energy barrier, so the system remains mainly around its minima
during the time evolution, caused by thermal fluctuations, in
simulation. Therefore, the configurations corresponding to the
unstable region at the middle are rare, consequently the relative
error for these data is large.

As the next step we proceed to calculate the critical temperatures
and critical exponents for continuous phase transitions, using
finite-size scaling theory.

\subsection{Determination of $T_{c}$ and static critical exponents}

 According to the finite-size scaling theory \cite{barber}, the
scaling form for various thermodynamic quantities  such as
magnetization density, susceptibility and specific heat in zero
field are given by:

\begin{eqnarray}
\label{mag}m&\approx& L^{\beta/\nu}{\mathcal{M}}(tL^{1/\nu})\\
\label{kappa}\chi&\approx& L^{\gamma/\nu}{\mathcal{K}}(tL^{1/\nu})\\
\label{sh}c&\approx&
c_{\infty}(t)+L^{\alpha/\nu}{\mathcal{C}}(tL^{1/\nu}),
\end{eqnarray}
where $t=(T-T_{c})/T_{c}$ is the reduced temperature for a
sufficiently large system at a temperature $T$ close enough to
the infinite lattice critical point $T_{c}$, $L$ is the linear
size of the system and  $\alpha,\beta,\gamma,\delta$ are static
critical exponents. Equations (\ref{mag}-\ref{sh}) are used to
estimate the critical exponents. However, before dealing with the
critical exponents  we should first determine the critical
temperature accurately.

The logarithmic derivatives of total magnetization ($mL^{d}$) are
important thermodynamic quantities for studying critical
phenomena and very useful to high accurate estimation of the
critical temperature $T_{c}$ and the correlation length critical
exponent ($\nu$)) \cite{chen}. To this, we Define the following
quantities:

\begin{eqnarray}
\label{v1}V_{1}&\equiv& 4[M^{3}]-3[M^{4}],\\
V_{2}&\equiv& 2[M^{2}]-[M^{4}],\\
V_{3}&\equiv& 3[M^{2}]-2[M^{3}],\\
V_{4}&\equiv& (4[M]-[M^{4}])/3,\\
V_{5}&\equiv& (3[M]-[M^{3}]/2,\\
\label{v6}V_{6}&\equiv& 2[M]-[M^{2}],
\end{eqnarray}
where $M=Nm$ is the total magnetization of the system and

\begin{equation}
[M^{n}]\equiv \ln\frac{\partial\langle M^{n} \rangle}{\partial T}.
\end{equation}
From Eq.(\ref{mag}) it is easy to show that

\begin{equation}
\label{vj}V_{j}\approx (1/\nu)\ln L+{\mathcal{V}}_{j}(tL^{1/\nu}),
\end{equation}
for $j=1,2,\cdot\cdot\cdot,6$. At the critical temperature
($t=0$), ${\mathcal{V}}_{j}$ should be constants, independent of
the system size $L$. Using Eq. (\ref{vj}) one can find the slope
of quantities $V_{1}$ to $V_{6}$ (Eq. \ref{v1}-\ref{v6}) versus
$\ln(L)$ for the region near the critical point. Scanning over
the critical region and looking for a quantity-independent slope
gives us both the critical temperature $T_{c}$ and the
correlation length exponent $\nu$ with high precision. Figures
(12) and (13) give the  examples of such an effort  for the set
of the coupling $D/|J|=0.2$. From these figures, we estimate that
$\nu=0.842(2)$ and $T_{c}=1.198(1)$. The linear fits to the data
in Fig.(12) have been obtained by the linear least squares method.

Once $\nu$ and $T_{c}$ are determined  accurately, we can extract
other static critical exponents related to the order parameter
($\beta$) and susceptibility ($\gamma$). The ratio $\beta/\nu$
can be estimated by using the size dependence of the order
parameter at the critical point given by Eq.(\ref{mag}).
Fig.(14)  shows the  log-log plots of the size dependence of the
order parameter corresponding to $D/|J|=0.5$ and $D/|J|=0.2$.
From this figure  the ratio $\beta/\nu$ can be estimated as the
slope of the straight lines fitted to the data according to
Eq.(\ref{mag}). We then have $\beta/\nu=0.198(8)$ for $D/|J|=0.5$
and $\beta/\nu=0.285(8)$ for $D/|J|=0.2$.

Accordingly, from Eq.(\ref{kappa}) it is clear that the peak
values of the finite-lattice susceptibility ($\chi=(\langle M^2
\rangle-\langle M \rangle^{2})/(NT)$) and the magnitude of the
true susceptibility at $T_{c}$ (the same as $\chi$ with $\langle m
\rangle=0$)  are asymptotically proportional to $L^{\gamma/\nu}$.
Then the slope of straight line fitted linearly to the log-log
plot of these two quantities versus linear size of the lattices,
can be calculated to estimate the ratio $\gamma/\nu$.  In Fig.(15)
the finite lattice susceptibility have been depicted for
${D/|J|}=0.2,0.5$, respectively. The slopes of linear lines
fitted to these data give  $\gamma/\nu=1.39(2)$ for ${D/|J|}=0.5$
and $\gamma/\nu=1.42(2)$ for ${D/|J|}=0.2$, where the error
includes the uncertainty in the slope resulting from uncertainty
in our estimate for $T_{c}$.

The above procedure has been applied for other values of
$D/|J|=1.0,0.5,0.2$ and the obtained critical exponents are listed
in Table.(II). In this table, the critical exponent $\alpha$, has
been calculated using the hyper-scaling relation:
\begin{equation}
\alpha=2-d\nu,
\end{equation}
in which $d=2$. On the other hand the Rushbrook scaling law
($\alpha+2\beta+\gamma=2$) is satisfied for all set of exponents
within the computational errors. For comparison, we have listed
the corresponding critical exponents of  Onsager's solution for
2D-Ising, and also Zamolodchikov's conjecture for the
Ising-tricritical point in two-dimensions, which corresponds to a
2D-$\phi^6$ field theory~\cite{tri1}. Zamolodchikov's conjecture
is based on conformal field theory and has been verified by Monte
Carlo simulation\cite{tri2}.

One can see from  Table.(II) that the critical exponents for
$D/|J|=1.0$ are pretty close to the 2D-Ising values, then
anisotropy magnitude of  $D/|J|=1.0$ is large enough to suppress
the transverse fluctuations normal to easy-axes directions. Upon
decreasing the anisotropy, the transverse fluctuations become
important and the exponents deviate from Ising values. However,
although the exponents $\nu$, $\gamma$ and $\alpha$ monotonously
tend to the the 2D-triciritcal values, but the exponent $\beta$
gets farther from it. This discrepancy, might the sign  of a new
universality class, other than 2D-$\phi^6$ model.

At the end, we deal with the dependence of the transition
temperature to the anisotropy intensity. We have already
mentioned the method of obtaining the critical temperature for
the continuous transitions ($D~ > ~0.17$). For strongly enough
first order transitions  whose energy histograms are double
peaked ($D~\langle~ 0.15$), the finite size transition
temperatures ($T_{c}(L)$) , are determined as the temperature at
which the two peaks have equal heights. Once $T_{c}(L)$ for all
lattice sizes is obtained, the transition temperature in
thermodynamic limit can be extrapolated by the following scaling
relation:

\begin{equation}
T_{c}(L)=T_{c}(\infty)+BL^{-d},
\end{equation}
where $B$ is a constant and $d=2$. The resulting  transition
temperatures are listed in Table.(I). In Fig.(16), we have plotted
the transition temperature versus $D$ in logarithmic scale. This
linear log-log plot shows a power law relation between these to
quantities as:

\begin{equation}
T_{c}\propto D^{0.501(2)}.
\end{equation}
This result is in clear contrast with mean-field prediction of a
linear dependence of transition temperature  on the anisotropy
intensity $D$. This scaling behaviour can be explained by a simple
dimensional analysis. Assuming that both exchange interaction,
$J$, and anisotropy, $D$, are equally important in occurrence
phase transition in $XY$ Kagom\'{e} antiferromagnet. So the
thermal energy which balances the entropy and internal energy at
the transition point, must be proportional to a combination of
$J$ and $D$.  Accordingly, dimensional analysis requires
$K_{B}T_{c}\sim (|J|D)^{1\over 2}$, which leads us to
$T_{c}/|J|\sim (D/|J|)^{1\over 2}$.

\section{Conclusion }

In summary, using mean-field theory and the optimized Monte Carlo
simulation based on multi-histogram, we investigated the phase
transitions of the antiferromagnetic classical $XY$ model on a
two dimensional $\mathrm{Kagom\acute{e}}$ lattice with the
easy-axes single ion anisotropy. In the absence of anisotropy,
this system is highly frustrated and  no phase transition is
expected to occur at finite temperatures, except the
Kosterlitz-Thouless transition mentioned in Ref. \cite{chandra}.
Turning on the anisotropy, removes the degeneracies of the ground
state and so establishes a long range order with all-in all-out
spin configuration at low temperatures. By increasing the
temperature, the system exhibits a phase transition from all-in
all-out ordered state to disordered (paramagnetic) state.
According to Monte Carlo results this transition is first order
for small values of anisotropy, while turns  to second order at a
tricritical point, corresponding to an anisotropy strength in the
interval $0.15 < \frac{D}{|J|} < 0.17$.

Employing finite size scaling theory, we derived the critical
exponents for continuous transitions and found that the transition
is in Ising universality for large values of anisotropy. This is
because in large $D/|J|$ limit, the fluctuations  perpendicular
to easy-axes directions are frozen, and so the effective degrees
of freedom  are spin flips along easy-axes directions,   such that
the order parameter possess the discrete $Z_{2}$ symmetry.
Decreasing the anisotropy magnitude, activates the spin
fluctuations perpendicular to the easy-axes directions. In
principle, the coupling of transverse modes (independent of
anisotropy) and also of other underlying modes, shown in Fig.(2)
and (3), with the all-in all-out state at $q=0$, is the reason
for the deviation of  the universality class  of transitions from
Ising, and is also responsible for changing the type of
transition to dis-continuous for small values of anisotropy.
However, obtained critical exponents near the tricritical point,
do not coincide with those of two-dimensional Ising-tricritical
point derived from 2D-$\phi^6$ field theory. This suggests the
possibility of the existence of a new tricritical universality
class in two-dimensions. It is not surprising, because the
critical behaviours in frustrated systems are usually different
form standard universality classes~\cite{kawamura2}. In this case,
finding such a universality class requires more theoretical and
numerical investigations.

We hope that this work  will motivate further experimental,
computational and analytical efforts for deeper understanding of
the nature of transitions in  geometrically frustrated systems.

{\textbf{ Acknowledgment}} \\
We would like to thank  M. J. P. Gingras, H. Kawamura, and P.
Holdsworth for enthusiastic discussions and useful comments.

%%%%%%%%%%%%%%%%%%%%%%%%%%%%%%%%%%%%%%%%%%%%%%%%%%%%%%%%%%%%%%%%%%
\begin{table}[c]
\begin{tabular}{|c|c|c|} %\hline
$D/J$ &  $T_{c}$ & $U^*$ \\
 \hline
 1.0     &  0.449(1)    &  0.66662(7)    \\
 0.5     &  0.316(1)    &  0.66660(9)             \\
 0.2     &  0.199(1)    &  0.66659(8)  \\
 0.18    &  0.189(5)    &  0.66653(9)  \\
 0.17    &  0.184(6)    &  0.66649(9)  \\
 0.15    &  0.174(8)    &  0.6664(1)  \\
 0.14    &  0.167(7)    &  0.6662(1)   \\
 0.13    &  0.162(7)    &  0.6661(1)   \\
 0.12    &  0.156(8)    &  0.6659(1)   \\
 0.1     &  0.142(8)    &  0.6658(1)
 %\hline
\end{tabular}
%{\box1}
\narrowtext\caption{The critical temperatures and value of $U^*$
for ${D\over
J}=1.0,0.5,0.2,0.18,0.17,0.15,0.14,0.13,0.12,0.1$.(see the text) }
\end{table}
%%%%%%%%%%%%%%%%%%%%%%%%%%%%%%%%%%%%%%%%%%%%%%%%%%%%%%%%%%%%%%%%%%%%%%%%%%%

%%%%%%%%%%%%%%%%%%%%%%%%%%%%%%%%%%%%%%%%%%%%%%%%%%%%%%%%%%%%%%%%%%
\begin{table}[t]
\begin{tabular}{|c|c|c|c|c|c|} %\hline
$D/|J|$  & $\nu$ & $\beta$ & $\gamma$ & $\alpha$ & $\alpha+2\beta+\gamma$\\
\hline
 1       &  1.019(2) & 0.15(1)  &  1.64(8) & -0.038(4)& 1.9(1)\\
 0.5     &  0.959(2) & 0.19(1)  &  1.52(6) & 0.082(4) & 2.0(1)\\
 0.2     &  0.842(2) & 0.24(2)  &  1.18(6) & 0.316(4) & 2.0(1)\\
 \hline
 2D-Ising         &  1 & 1/8  &  7/4 & 0($\log$) &  2\\
 2D-$\phi^6$   &  5/9 & 1/24  &  37/36 & 8/9 &   2\\
\end{tabular}
%{\box1}
\narrowtext\caption{The  static critical exponents $\nu, \beta,
\gamma$ and $\alpha$ for ${D\over J}=1.0,0.5,0.2$, derived from
finite-size scaling. In the last column the Rushbrook's scaling
law is computed. The  last two rows are listed the  corresponding
exact critical exponent of 2D-Ising model and two-dimensional
Ising-tricriticl point, respectively.}
\end{table}
%%%%%%%%%%%%%%%%%%%%%%%%%%%%%%%%%%%%%%%%%%%%%%%%%%%%%%%%%%%%%%%%%%%%%%%%%%%

%1
\newpage
%%%%%%%%%%%%%%%%%%%%%%%%%%%%%%%%%%%%%%%%%%%%%%%%%%%%%%%%%%%%%%%%%%%%%%%%%
\begin{figure}[c]
{\epsfxsize=10.5truecm  \epsfbox{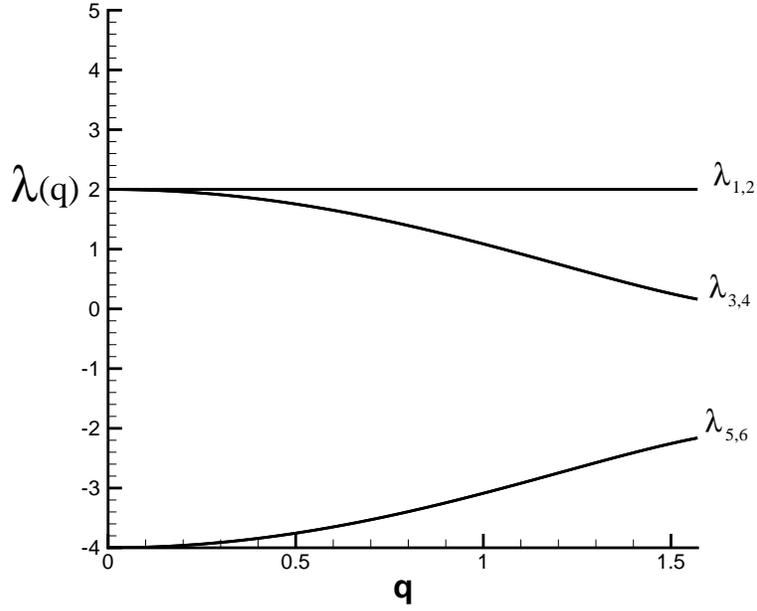}}
\vspace{1cm}\narrowtext \caption{ spectrum of coupling matrix
${\tilde J}$ for $D=0$ along [10] direction. Each branch has two
fold degeneracy. }\label {1a}
\end{figure}
%%%%%%%%%%%%%%%%%%%%%%%%%%%%%%%%%%%%%%%%%%%%%%%%%%%%%%%%%%%%%%%%%%%%%%%%%%%
%2
\newpage
%%%%%%%%%%%%%%%%%%%%%%%%%%%%%%%%%%%%%%%%%%%%%%%%%%%%%%%%%%%%%%%%%%%%%%%%%
\begin{figure}[c]
{\epsfxsize=10.5truecm  \epsfbox{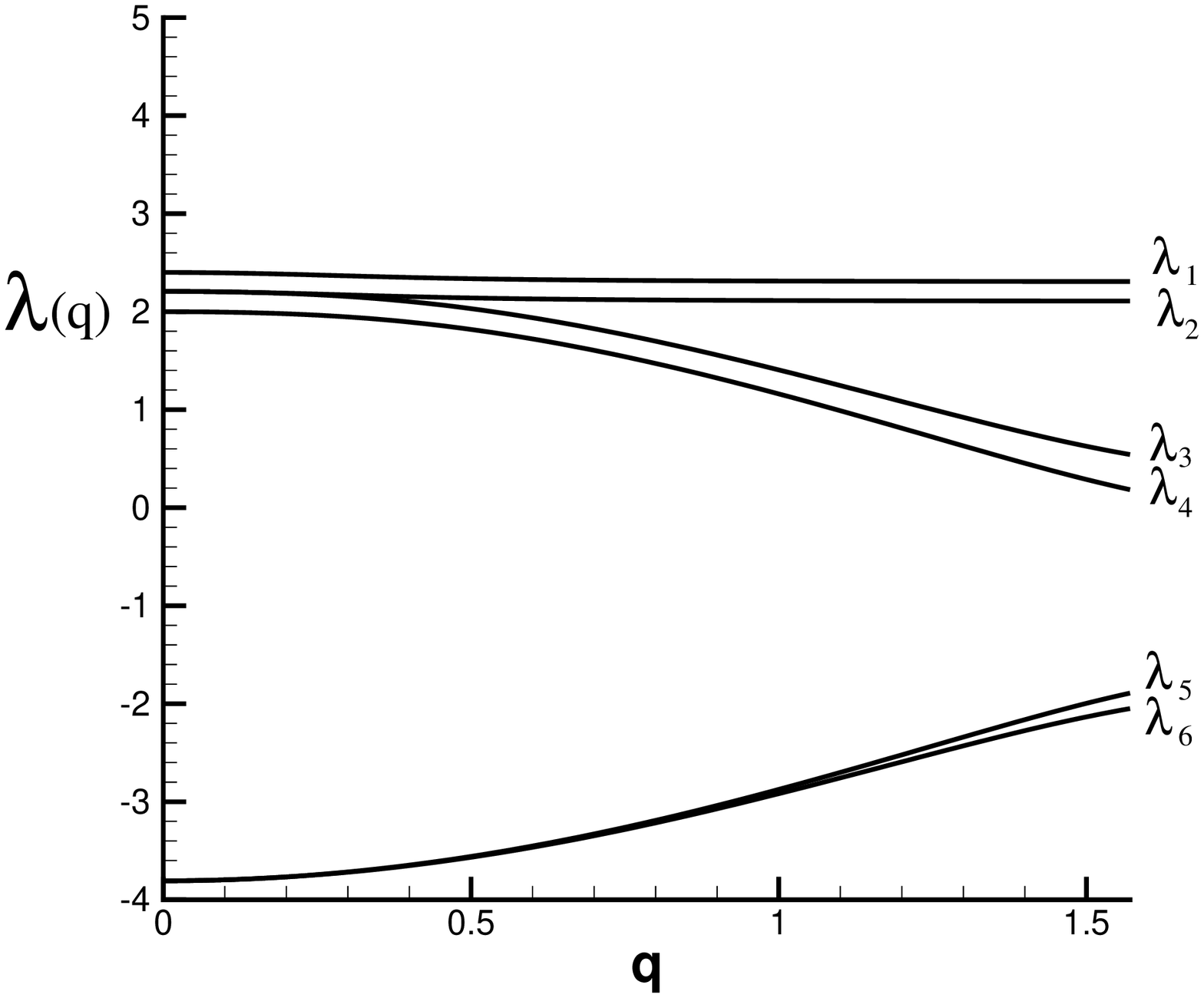}}
\vspace{1cm}\narrowtext \caption{ spectrum of coupling matrix
${\tilde J}$ for $D=0.2$ along [10] direction. degeneracies have
been removed by addition anisotropic term. }\label{2a}
\end{figure}
%%%%%%%%%%%%%%%%%%%%%%%%%%%%%%%%%%%%%%%%%%%%%%%%%%%%%%%%%%%%%%%%%%%%%%%%%%%
%3
\newpage
%%%%%%%%%%%%%%%%%%%%%%%%%%%%%%%%%%%%%%%%%%%%%%%%%%%%%%%%%%%%%%%%%%%%%%%%%
\begin{figure}[c]
{\epsfxsize=10.5truecm  \epsfbox{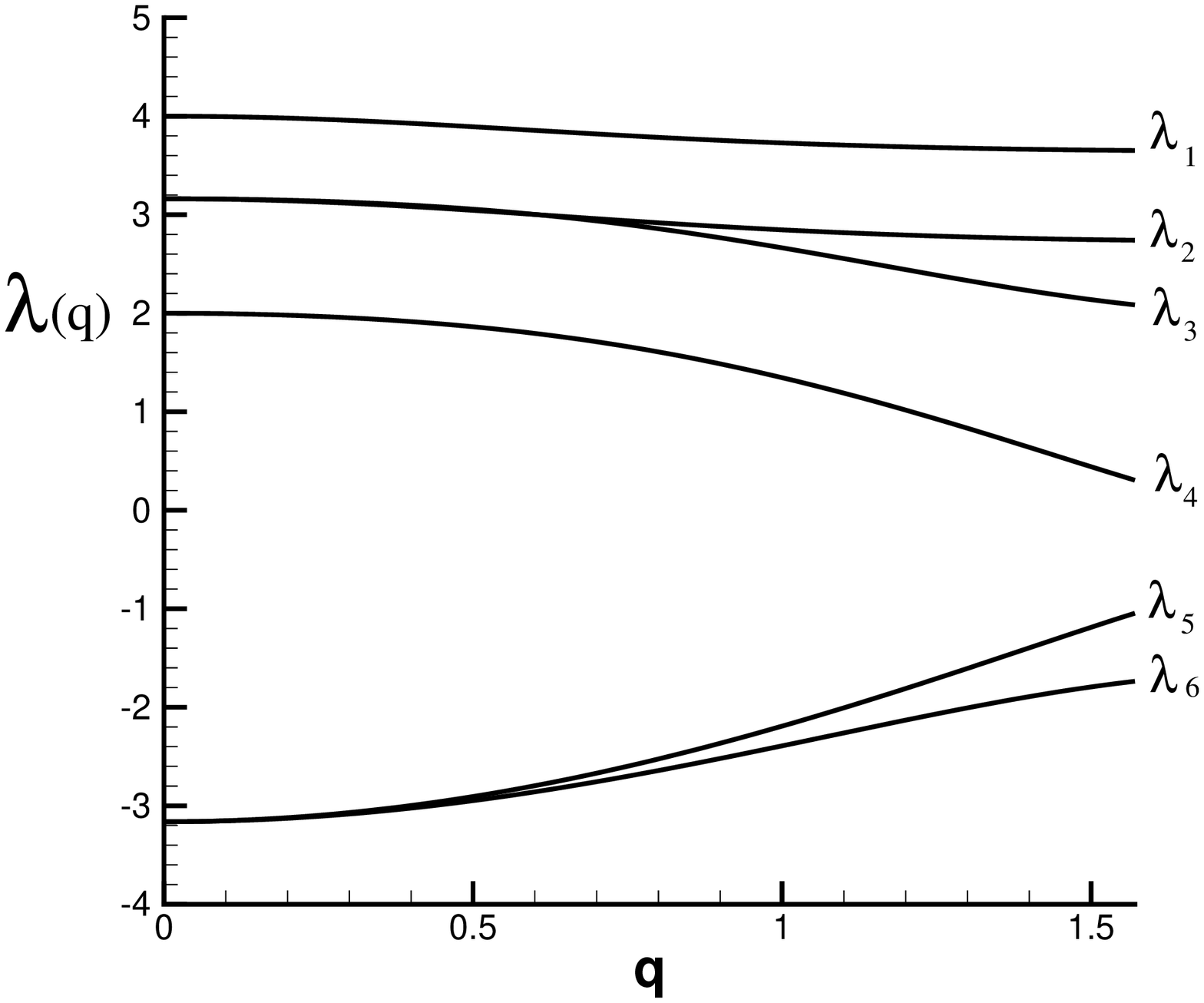}}
\vspace{1cm}\narrowtext \caption{ spectrum of coupling matrix
${\tilde J}$ for $D=1.0$ along [10] direction. degeneracies have
been removed by addition anisotropic term. }\label{3a}
\end{figure}
%%%%%%%%%%%%%%%%%%%%%%%%%%%%%%%%%%%%%%%%%%%%%%%%%%%%%%%%%%%%%%%%%%%%%%%%%%%
%4
\newpage
%%%%%%%%%%%%%%%%%%%%%%%%%%%%%%%%%%%%%%%%%%%%%%%%%%%%%%%%%%%%%%%%%%%%%%%%%
\begin{figure}[c]
\centerline{\epsfxsize=10.5truecm  \epsfbox{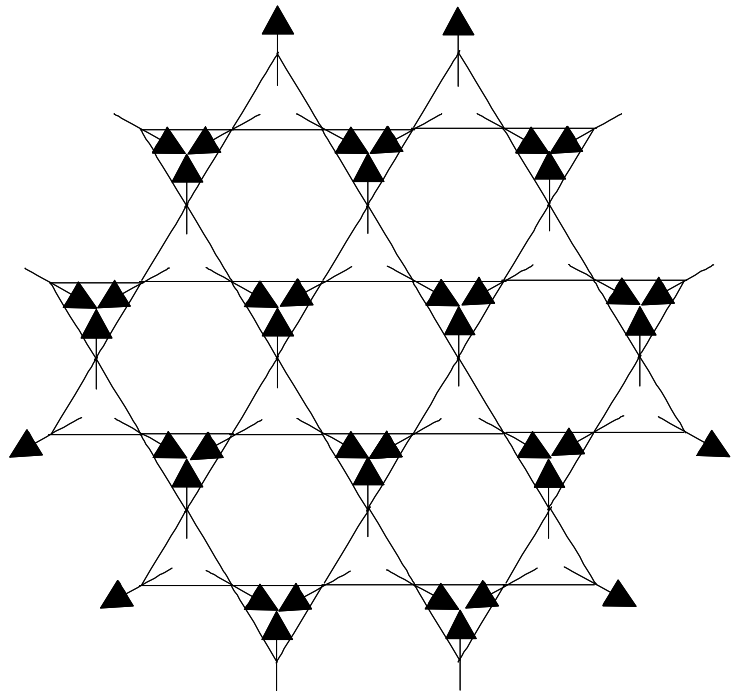}} \narrowtext
\caption{ All in-all out configuration in kagome' lattice.
}\label{4a}
\end{figure}
%%%%%%%%%%%%%%%%%%%%%%%%%%%%%%%%%%%%%%%%%%%%%%%%%%%%%%%%%%%%%%%%%%%%%%%%%%%

%5
\newpage
%%%%%%%%%%%%%%%%%%%%%%%%%%%%%%%%%%%%%%%%%%%%%%%%%%%%%%%%%%%%%%%%%%%%%%%%%
\begin{figure}[c]
{\epsfxsize=10.5truecm \epsfbox{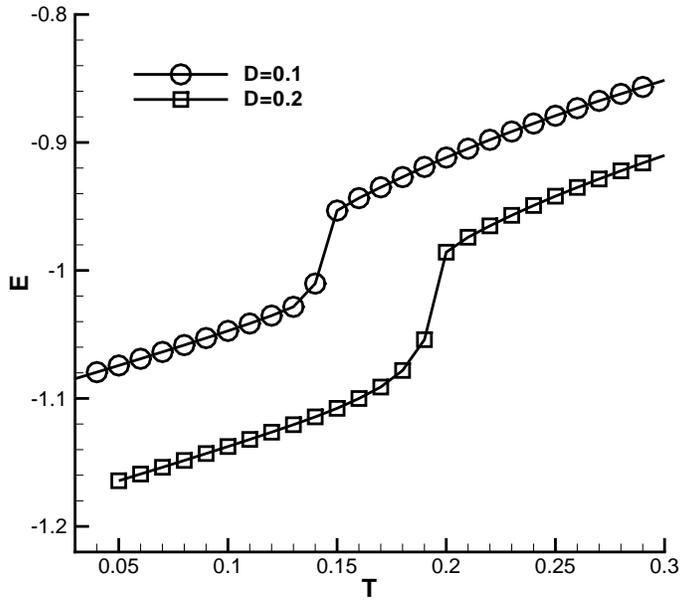}} \vspace{1cm}\narrowtext
\caption{ Temperature dependence of Energy per spin for
$D=0.2,0.1$.}
\end{figure}
%%%%%%%%%%%%%%%%%%%%%%%%%%%%%%%%%%%%%%%%%%%%%%%%%%%%%%%%%%%%%%%%%%%%%%%%%%%

%6
\newpage
%%%%%%%%%%%%%%%%%%%%%%%%%%%%%%%%%%%%%%%%%%%%%%%%%%%%%%%%%%%%%%%%%%%%%%%%%
\begin{figure}[c]
{\epsfxsize=10.5truecm \epsfbox{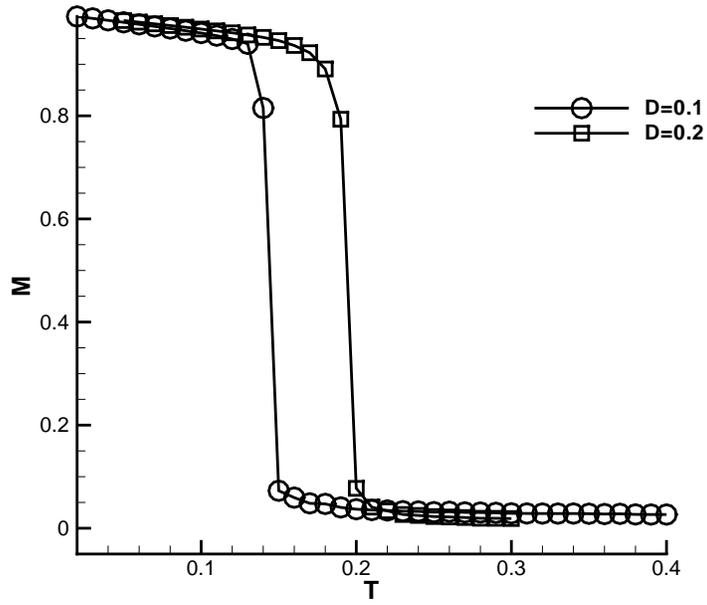}} \vspace{1cm}\narrowtext
\caption{ Temperature dependence  of order
parameter(magnetization) for $D=0.2,0.1$.}
\end{figure}
%%%%%%%%%%%%%%%%%%%%%%%%%%%%%%%%%%%%%%%%%%%%%%%%%%%%%%%%%%%%%%%%%%%%%%%%%%%

%7
\newpage
%%%%%%%%%%%%%%%%%%%%%%%%%%%%%%%%%%%%%%%%%%%%%%%%%%%%%%%%%%%%%%%%%%%%%%%%%
\begin{figure}[c]
{\epsfxsize=10.5truecm \epsfbox{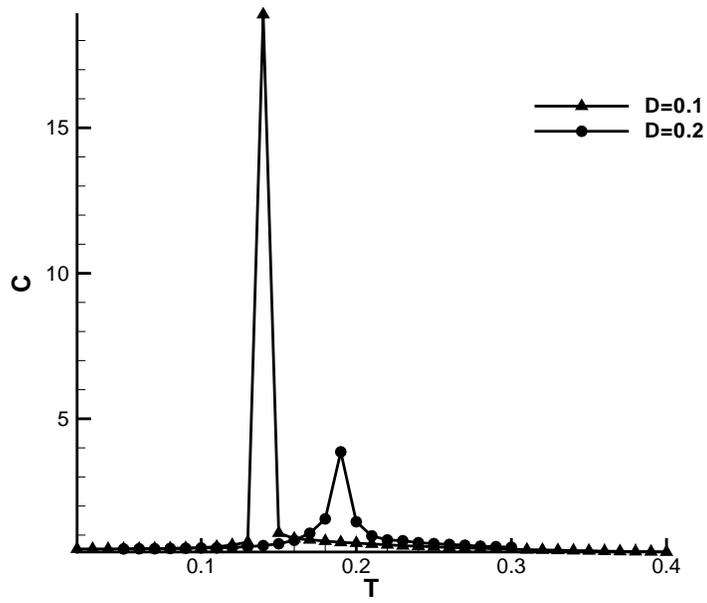}} \vspace{1cm}\narrowtext
\caption{ Temperature dependence  of specific heat for
$D=0.2,0.1$.}
\end{figure}
%%%%%%%%%%%%%%%%%%%%%%%%%%%%%%%%%%%%%%%%%%%%%%%%%%%%%%%%%%%%%%%%%%%%%%%%%%%

%8
\newpage
%%%%%%%%%%%%%%%%%%%%%%%%%%%%%%%%%%%%%%%%%%%%%%%%%%%%%%%%%%%%%%%%%%%%%%%%%
\begin{figure}[c]
{\epsfxsize=10.5truecm \epsfbox{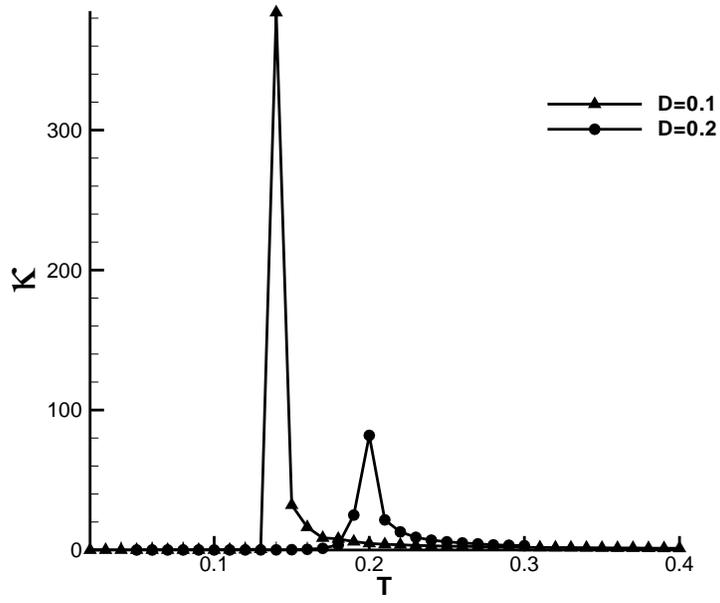}} \vspace{1cm}\narrowtext
\caption{Temperature dependence  of susceptibility for
$D=0.1,0.2$.}
\end{figure}
%%%%%%%%%%%%%%%%%%%%%%%%%%%%%%%%%%%%%%%%%%%%%%%%%%%%%%%%%%%%%%%%%%%%%%%%%%%

%9
\newpage
%%%%%%%%%%%%%%%%%%%%%%%%%%%%%%%%%%%%%%%%%%%%%%%%%%%%%%%%%%%%%%%%%%%%%%%%%
\begin{figure}[c]
{\epsfxsize=10.5truecm \epsfbox{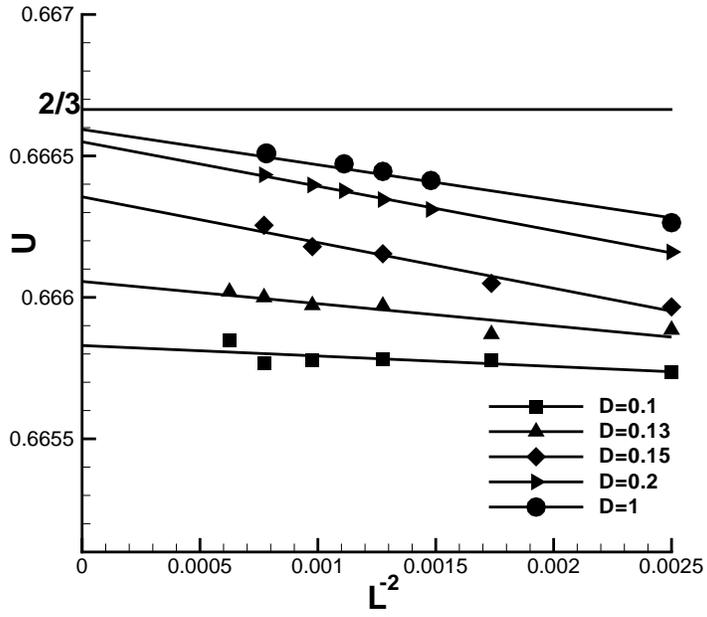}} \vspace{1cm}\narrowtext
\caption{ Size dependences  of binder's fourth energy cumulant for
$D=1.0,0.2,0.15,0.13,0.1$.}
\end{figure}
%%%%%%%%%%%%%%%%%%%%%%%%%%%%%%%%%%%%%%%%%%%%%%%%%%%%%%%%%%%%%%%%%%%%%%%%%%%

%10
\newpage
%%%%%%%%%%%%%%%%%%%%%%%%%%%%%%%%%%%%%%%%%%%%%%%%%%%%%%%%%%%%%%%%%%%%%%%%%
\begin{figure}[c]
{\epsfxsize=10.5truecm\epsfbox{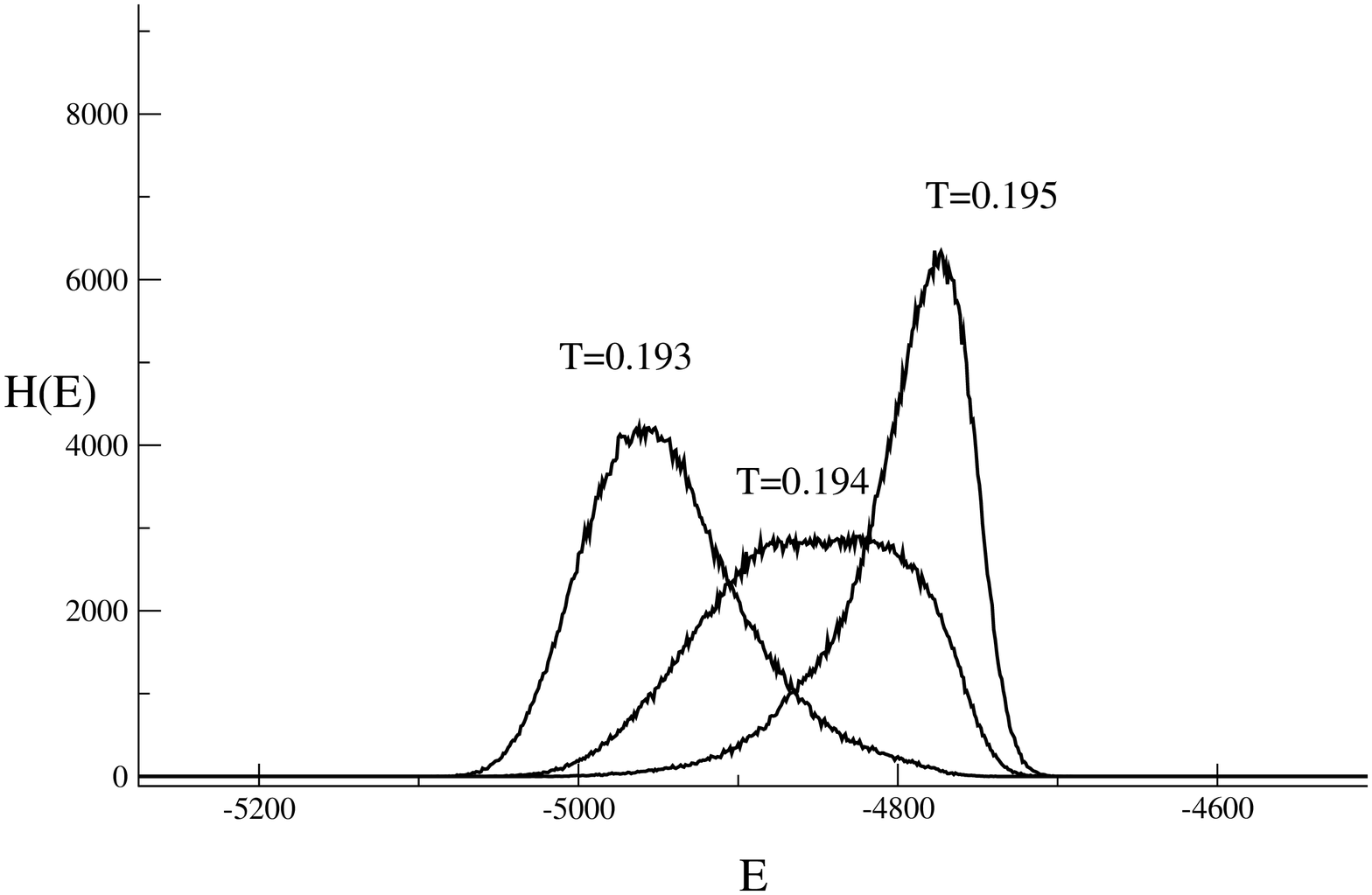}}\vspace{1cm}\narrowtext\caption{Three
energy histograms  for $D=0.2$ and size $N=3\times 40\times 40$
near the transition temperature.}
\end{figure}
%%%%%%%%%%%%%%%%%%%%%%%%%%%%%%%%%%%%%%%%%%%%%%%%%%%%%%%%%%%%%%%%%%%%%%%%%%%

%11
\newpage
%%%%%%%%%%%%%%%%%%%%%%%%%%%%%%%%%%%%%%%%%%%%%%%%%%%%%%%%%%%%%%%%%%%%%%%%%
\begin{figure}[c]
{\epsfxsize=10.5truecm \epsfbox{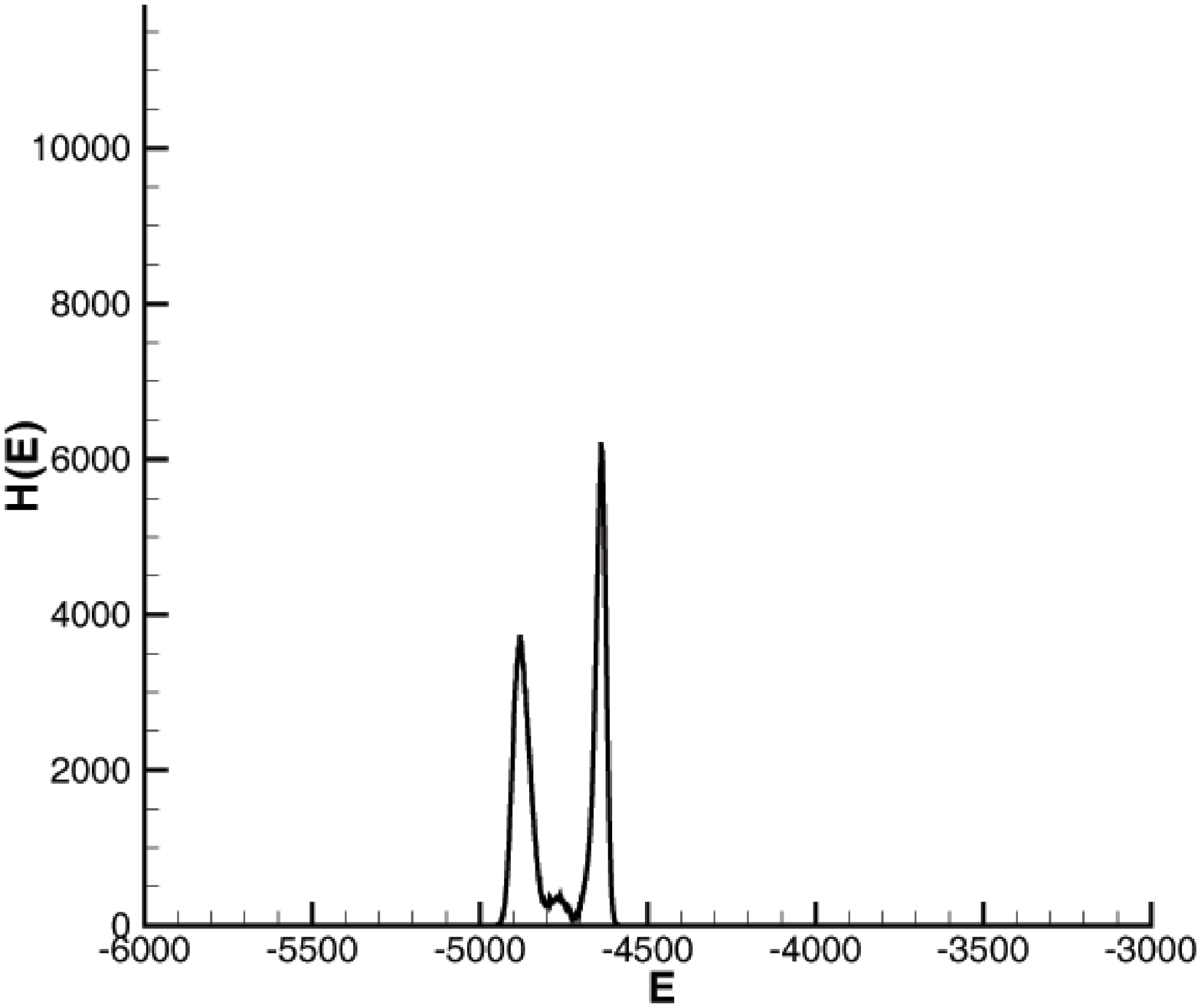}}
\vspace{1cm}\narrowtext \caption{Energy histogram  for $D=0.1$
and size $N=3\times 40\times 40$ near the transition temperature.}
\end{figure}
%%%%%%%%%%%%%%%%%%%%%%%%%%%%%%%%%%%%%%%%%%%%%%%%%%%%%%%%%%%%%%%%%%%%%%%%%%%

%12
\newpage
%%%%%%%%%%%%%%%%%%%%%%%%%%%%%%%%%%%%%%%%%%%%%%%%%%%%%%%%%%%%%%%%%%%%%%%%%
\begin{figure}[c]
{\epsfxsize=10.5truecm \epsfbox{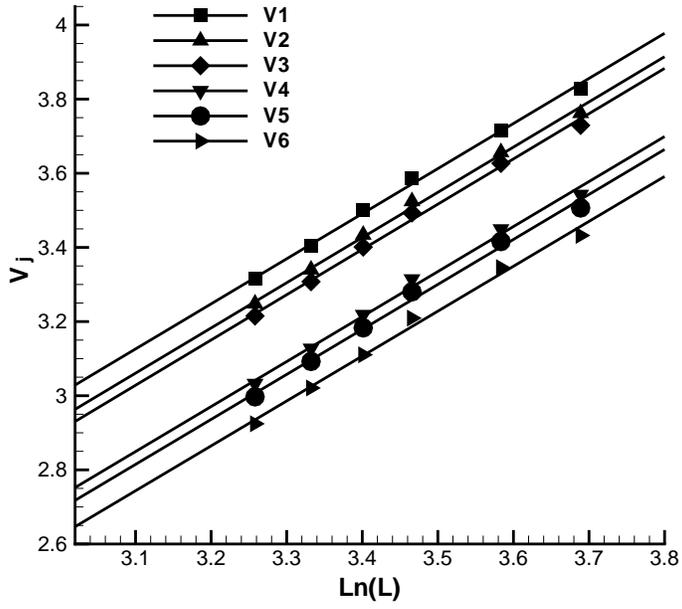}}
\vspace{1cm}\narrowtext \caption{ Dependence of quantity $V_{j}$
(see the text) versus logarithm of $L$  for $D=0.2$ at
$T=0.1989(5)$. The solid lines represent linear fits to
Eq.(\ref{vj}). All straight lines have the same slope
$\nu=0.842(2)$.}
\end{figure}
%%%%%%%%%%%%%%%%%%%%%%%%%%%%%%%%%%%%%%%%%%%%%%%%%%%%%%%%%%%%%%%%%%%%%%%%%%%

%13
\newpage
%%%%%%%%%%%%%%%%%%%%%%%%%%%%%%%%%%%%%%%%%%%%%%%%%%%%%%%%%%%%%%%%%%%%%%%%%
\begin{figure}[c]
{\epsfxsize=10.5truecm \epsfbox{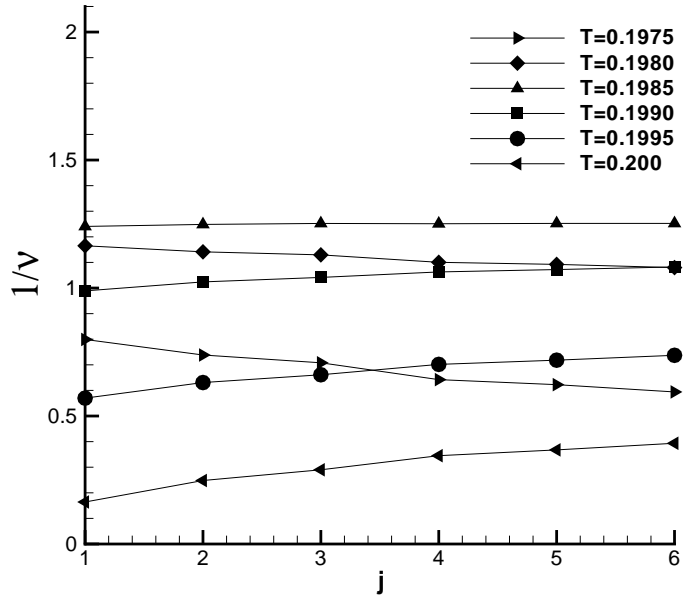}}
\vspace{1cm}\narrowtext \caption{ Scanning results for the
dependence of quantity $V_{j}$  versus $j$ for $D=0.2$. The
horizontal line is drawn at $1/\nu=1.187$.}
\end{figure}
%%%%%%%%%%%%%%%%%%%%%%%%%%%%%%%%%%%%%%%%%%%%%%%%%%%%%%%%%%%%%%%%%%%%%%%%%%%

%14
\newpage
%%%%%%%%%%%%%%%%%%%%%%%%%%%%%%%%%%%%%%%%%%%%%%%%%%%%%%%%%%%%%%%%%%%%%%%%%
\begin{figure}[c]
{\epsfxsize=10.5truecm \epsfbox{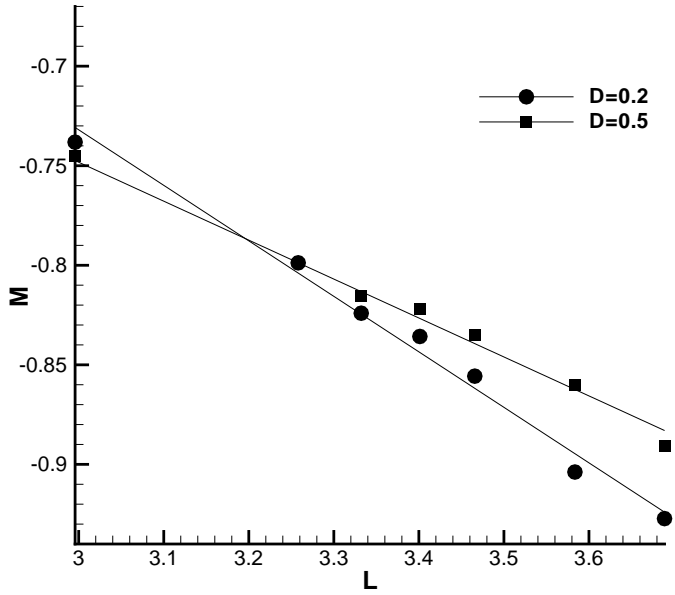}}
\vspace{1cm}\narrowtext \caption{ Log-Log plot of order parameter
for $D=0.5,0.2$. The slopes of fitted line gives
$\frac{\beta}{\nu}=0.285(7)$ for $D=0.2$ and
$\frac{\beta}{\nu}=0.198(8)$ and for $D=0.5$.}
\end{figure}
%%%%%%%%%%%%%%%%%%%%%%%%%%%%%%%%%%%%%%%%%%%%%%%%%%%%%%%%%%%%%%%%%%%%%%%%%%%

%15
\newpage
%%%%%%%%%%%%%%%%%%%%%%%%%%%%%%%%%%%%%%%%%%%%%%%%%%%%%%%%%%%%%%%%%%%%%%%%%
\begin{figure}[c]
{\epsfxsize=10.5truecm \epsfbox{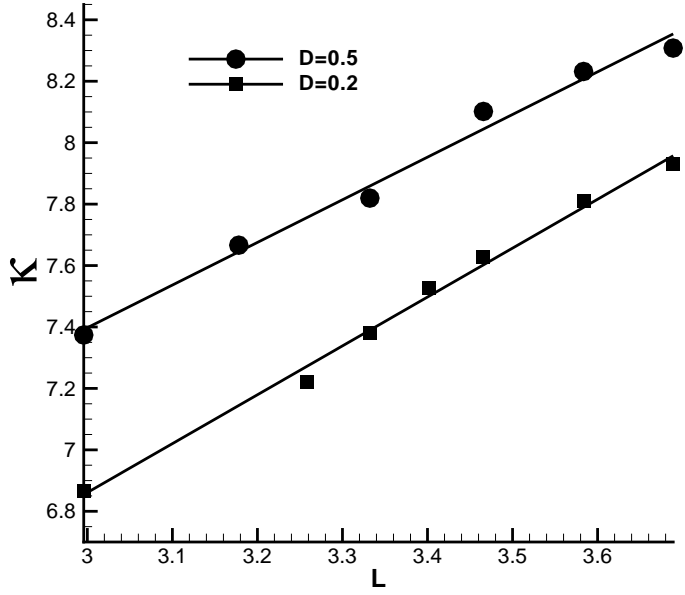}}
\vspace{1cm}\narrowtext \caption{ Log-Log plot of finite lattice
susceptibility for $D=0.2,0.5$. The slopes of fitted lines give
$\frac{\gamma}{\nu}=1.58(5)$ for $D=0.2$ and
$\frac{\gamma}{\nu}=1.40(5)$ for one $D=0.5$.}
\end{figure}
%%%%%%%%%%%%%%%%%%%%%%%%%%%%%%%%%%%%%%%%%%%%%%%%%%%%%%%%%%%%%%%%%%%%%%%%%%%

%16
\newpage
%%%%%%%%%%%%%%%%%%%%%%%%%%%%%%%%%%%%%%%%%%%%%%%%%%%%%%%%%%%%%%%%%%%%%%%%%
\begin{figure}[c]
{\epsfxsize=10.5truecm \epsfbox{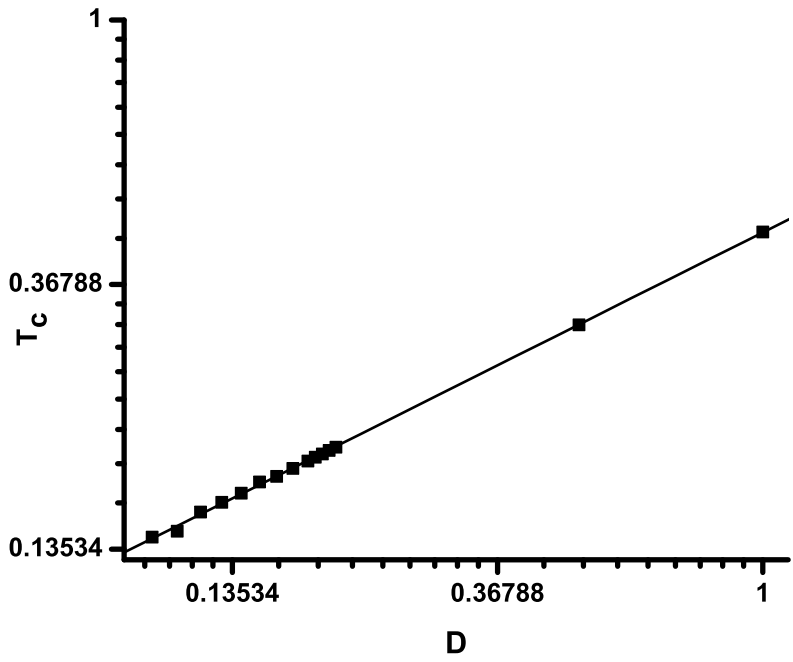}}
\vspace{1cm}\narrowtext \caption{ Log-Log plot of transition
temperature versus anisotropy magnitude $D$. The slope of fitted
line is $0.501(2)$. }
\end{figure}
%%%%%%%%%%%%%%%%%%%%%%%%%%%%%%%%%%%%%%%%%%%%%%%%%%%%%%%%%%%%%%%%%%%%%%%%%%%

%\newpage
%%%%%%%%%%%%%%%%%%%%%%%%%%%%%%%%%%%%%%%%%%%%%%%%%%%%%%%%%%%%%%%%%%%%%%%%%
%\begin{figure}[c]
%{\epsfxsize=10.5truecm \epsfbox{fig18.eps}}
%\vspace{1cm}\narrowtext \caption{ Dependence of $T_C$ in $L^{-2}$
%scale for $D=0.1$. Linear fit show that in $L^{-2}\rightarrow 0$
%critical temperature reach to $T_c=0.142(1)$. }
%\end{figure}
%%%%%%%%%%%%%%%%%%%%%%%%%%%%%%%%%%%%%%%%%%%%%%%%%%%%%%%%%%%%%%%%%%%%%%%%%%%

\end{document}